\documentclass[aps,showpacs,preprintnumbers,amsmath,amssymb,nofootinbib]{revtex4}

\usepackage{subfigure}
\usepackage{graphicx}
\usepackage{color}
\usepackage{amsmath}
\usepackage{amsfonts}
\usepackage{amssymb}
\def \be  {\begin{equation}}
\def \ee  {\end{equation}}
\def \ee  {\end{equation}}
\def \bea {\begin{eqnarray}}
\def \eea {\end{eqnarray}}

\begin{document}

\preprint{ECTP-2019-06}
\preprint{WLCAPP-2019-06}

\title{Particle Ratios within EPOS, UrQMD and Thermal Models at AGS, SPS and RHIC Energies}

\author{Mahmoud Hanafy}
\email{mahmoud.nasar@fsc.bu.edu.eg}
\affiliation{Physics Department, Faculty of Science, Benha University, 13518, Benha, Egypt}
\affiliation{World Laboratory for Cosmology And Particle Physics (WLCAPP), 11571 Cairo, Egypt}

\author{Abdel Nasser Tawfik}
\email{atawfik@nu.edu.eg}
\affiliation{Nile University - Egyptian Center for Theoretical Physics (ECTP), Juhayna Square off 26th-July-Corridor, 12588 Giza, Egypt}
\affiliation{Institute for Theoretical Physics, Goethe University, Max-von-Laue-Str.  1, D-60438 Frankfurt am Main, Germany}

\author{Muhammad Maher}
\email{m.maher@science.helwan.edu.eg}
\affiliation{Helwan University, Faculty of Science, Physics Department, Ain Helwan, P.O. 11795, 11795 Cairo, Egypt}
\affiliation{World Laboratory for Cosmology And Particle Physics (WLCAPP), 11571 Cairo, Egypt}

\author{Werner Scheinast}
\email{werner@jinr.ru}
\affiliation{Joint Institute for Nuclear Research - Veksler and Baldin Laboratory of High Energy Physics, Moscow Region, 141980 Dubna, Russia}

\date{\today}

\begin{abstract}  
 		
The particle ratios $k^+/\pi^+$, $\pi^-/K^-$, $\bar{p}/\pi^-$, $\Lambda/\pi^-$, $\Omega/\pi^-$, $p/\pi^+$, $\pi^-/\pi^+$, $K^-/K^+$, $\bar{p}/p$, $\bar{\Lambda}/\Lambda$, $\bar{\Sigma}/\Sigma$, $ \bar{\Omega}/\Omega$ measured at AGS, SPS and RHIC energies are compared with large statistical ensembles of $100,000$ events deduced from the CRMC EPOS $1.99$ and the Ultra-relativistic Quantum Molecular Dynamics (UrQMD) hybrid model. In the UrQMD hybrid model two types of phase transitions are taken into account. All these are then confronted to the Hadron Resonance Gas Model.  The two types of phase transitions are apparently indistinguishable. Apart from $k^+/\pi^+$, $k^-/\pi^-$, $\Omega/\pi^-$,  $\bar{p}/\pi^+$, and $\bar{\Omega}/\Omega$, the UrQMD hybrid model agrees well with the CRMC EPOS $1.99$. Also, we conclude that the CRMC EPOS $1.99$ seems to largely underestimate  $k^+/\pi^+$, $k^-/\pi^-$, $\Omega/\pi^-$, and $\bar{p}/\pi^+$.

\end{abstract}. 


\maketitle


\section{Introduction}

In nuclear collisions, the statistical nature of the particle production allows the utilization of the particle ratios, for instance, to conduct systematic studies on the thermal properties at final state. Over the last few decades, huge experimental data at energies covering up four orders of magnitude of GeV are now available. It turns out that various statistical thermal models \cite{r2,r4,r5,r7,r8,r9,r10,r11,r12} are remarkably successful in explaining the resulting particle yields and their ratios measured in heavy-ion collisions. Such a huge data set allowed us to draw the conclusion that the produced particles seem reaffirming the assumption that the hadrons are likely stemming from thermal sources with given temperatures and given chemical potentials. It becomes obvious that such a thermal nature is valid, universally, except for a few baryon-to-meson ratios, such as proton-to-pion at top RHIC and LHC, known as proton anomaly \cite{Tawfik:2019wze}. 
 
Besides the statistical narture of the particle production, another main goal of the nuclear collisions is the detection of unampagious signatures for the possible hadron-quark phase transition \cite{r1}. This can, among others, allow us to verify the theory of strong interaction, the quantum chromodynamics (QCD), which predicts that the hadronic matter likely undergeos phase transition(s) from confined colorless hadrons to deconfined colored quark-gluon plasma (QGP) or vice versa \cite{r2}. So far various nuclear experiments gave indirect signatures for the existance of the QGP phase, for example ref. \cite{McLerran2004,r3}. So far, we understand that the order of the phase transition, especially at low baryon chemical potential, is a rapid crossover \cite{Bazavov:2018mes,Tawfik:2019rdd,Tawfik:2016gye}. 

It was argued that at equilibrium the particle ratios are well explained by two variables; the freezeout temperature ($T_{\mathtt{ch}}$) and the baryon chemical potential ($\mu_{\mathtt{b}}$). In nuclear collisions, the freezeout stage occurs where the inellastic reactions cease and the number of produced particles becomes fixed. At this stage, the thermal models, such as the hadron resonance gas (HRG) model, determine essential characteristics for dense and hot fireball generated in the heavy-ion collisions. As a result, the thermal models are utilized as an essential tool connecting the QCD phase diagram with the nuclear experiments \cite{rr12, rr13}, in the sense that the measurements, mainly the particle multiplicities are connected to the number density in the thermal models, which in turn are strongly depending on chemical freezeout parameters; $T_{\mathtt{ch}}$ and $\mu_{\mathtt{b}}$. In this way, we speak of thermodynamics, correlations, fluctuations, etc. in nuclear collisions \cite{r4}.
 
The dependenece $\mu_{\mathtt{b}}$ and $T_{\mathtt{ch}}$ on the nucleon-nulceon center-of-mass energies $\sqrt{s_{\mathtt{NN}}}$ contructs a boundary of the chemical freezeout diagram which is very close to the QCD phase diagram \cite {r18}. The dependence of $T_{\mathtt{ch}}$ on $\mu_{\mathtt{b}}$ looks similar to that of the various thermodynamic quantities as calculated in the lattice QCD \cite {r19,r20}, which in turn are relaible quantities, especially at $\mu_b/T\leq 1$, i.e. at $\sqrt{s_{\mathtt{NN}}}$ greater than that of top SPS energies. Accordingly, these boundries remain unfeatured \cite{Tawfik:2016jzk}, at lower energies, i.e. larger baryon chemical potential, where the  QCD-like effective models, such as the HRG model \cite{r4} and the Polyakov linear-sigma model (PLSM)  \cite{Tawfik:2019rdd,Tawfik:2016gye} play a major role. The hybrid event-generators such as the Cosmic Ray Monte Carlo (CRMC) \cite{q0149, q0150, qii51, qii52, qii53, S54, S55, S56, EE1, EE2} models and the Ultra-relativistic Quantum Molecular Dynamic (UrQMD) v$3.4$ \cite{r26,r27,r28,r29,r30} are the frameworks, which we are going to compare with the available experimental results and with calculations based on the thermal models, as well.
 
The present work presents predictions for the future facilities such the Nuclotron-based Ion Collider fAcility (NICA) future facility at the Joint Institute for Nuclear Research (JINR), Dubna-Russia and the Facility for Antiproton and Ion Research (FAIR) at the Gesellschaft f\"ur Schwerionenforsching (GSI), Darmstadt-Germany. These and the BES-II program at RHIC are designed to cover the intermediate temperature region of the QCD phase diagram, while both LHC and top RHIC obviously operate at low $\mu_{\mathtt{b}}$ or high $T_{\mathtt{ch}}$, i.e. left part of the QCD phase-diagram.    

In the present paper, we compare various particle ratios deduced from CRMC and UrQMD v$3.4$ with  the HRG calculations \cite{Tawfik:2016jzk}. The latter would allow us to adjust $T_{\mathtt{ch}}$ and $\mu_{\mathtt{b}}$, if we primarily were interested in statistical fits. In the present study, we aren't tageting any statistical fits. We concretly aim at comparing these three sets of results, namely the experimental results, the results deduced from the two event generators, and the HRG calculations. The latter use a combination of $T_{\mathtt{ch}}$ and $\mu_{\mathtt{b}}$  in order to deduce various partcile yields and ratios, at a wide range of energies. Here, we focus on enegrgies ranging from $7.7$ to $\sim 200~$GeV. In the HRG calculations, both $T_{\mathtt{ch}}$ and $\mu_{\mathtt{b}}$ are conditioned to one of the universal freezeout conditions \cite{Tawfik:2016jzk}, such as constant entropy density normalized to $T_{\mathtt{ch}}^3$ \cite{Tawfik:2005qn,Tawfik:2004ss}, constant higher moments of the particle multiplicity \cite{Tawfik:2013dba, Tawfik:2012si}, constant trace anomaly \cite{Tawfik:2013eua} or an analogy of the Hawking-Unruh radiation \cite{Tawfik:2015fda}. 

With this reference, we highlight that two types of the phase transitions are taken into account in UrQMD, namely, first order and crossover. We emphasize that the HRG model, which is a good statistical approach of various thermodynamic quantities, such as the particle density, can't be utilized for phenomena like deconfinement and chiral phase transition(s) \cite{Tawfik:2016jzk}. But HRG describes well the hadron phase \cite{r4}. Different stages of the colliding systems covering from early stages up to the final state of the particle production can be characterized in UrQMD v3.4  \cite{r31} and in CRMC \cite{q0149}. Out of the various types of the nucleus-nucleus collisions in CRMC, we utilize the hadronic interacting model namely EPOS $1.99$. Having all these prepared, a direct comparison with the experimental results can be achieved.

The present paper is organized as follows. Section \ref{models} gives short reviews on the different approaches; HRG, UrQMD v$3.4$ and CRMC EPOS $1.99$. In section \ref{results}, the results on different particle ratios are presented. Section \ref{conc} is devoted to the final conclusions.
	
\section{Approaches and Event Generators}
\label{models}

In this section, we give a short description on the Cosmic Ray Montocarlo Code (CRMC EPOS $1.99$) and the Ultrarelativistic Quantum Molecular Dynamic (UrQMD) hybrid approaches which shall be used in calculating various particle ratios at energies spanning between $\sqrt{s_{NN}}=7.7$ and $200~$GeV. The comparison between the results from CRMC and UrQMD v$3.4$ event generators and that from the HRG model with AGS, SPS and RHIC experiment is novel on one hand side. On the other hand, this allows us to conduct a systematic study. We aim at understanding whether both even generators CRMC and UrQMD v$3.4$ are able to give particle ratios compatible to the experiments and accordingly shed light on dynamics of the particle production and how this would be depending on the beam energy. Furthermore, this would help in validating both event generators in estimating different particle ratios and simultaneously deducing the freezeout parameters, especially at energies where experimental results aren't available so far and/or where FAIR and NICA shall be operating.

\subsection{Hadron resonance gas (HRG) model} 
\label{sec:hrg}  
  
As per Hagedorn, the formation of resonances is to be understood in a bootstrap picture, i.e. the resonances or fireballs are conjectured being composed of further resonances or fireballs, which in turn are consistent of lighter resonances or smaller fireballs and so on. The thermodynamic properties of such a system can be derived directly from the partition function $Z(T, \mu, V)$. In a grand canonical ensemble, this reads \cite{r4}
\bea 
Z(T,V,\mu)=\mbox{Tr}\left[\exp\left(\frac{{\mu}N-H}{T}\right)\right], \label{GrindEQ__1_}
\eea
where $H$ is Hamiltonian combining all relevant degrees of freedom (dof) in deconfined and strongly interacting system and $N$ is the number of constituents, e.g. dof. In the HRG model, Eq. (\ref{GrindEQ__1_}) can be expressed as a sum over all hadron resonances \cite{r4}. These can be taken from the recent particle data group \cite{pdg}
\begin{equation} 
\ln  Z(T,V,\mu)=\sum_i{{\ln Z}_i(T,V,\mu)} =\frac{V g_i}{2{\pi}^2}\int^{\infty}_0{\pm p^2 dp {\ln} {\left[1\pm {\lambda}_i \exp\left(\frac{-{\varepsilon}_i(p)}{T} \right) \right]}}, \label{GrindEQ__2_}
\end{equation}
where $\pm$ stands for bosons and fermions, respectively, $\varepsilon_{i}=\left(p^{2}+m_{i}^{2}\right)^{1/2}$ is the dispersion relation of the $i$-th particle and $\lambda_i$ is its fugacity factor \cite{r4}
\begin{equation} \label{GrindEQ__4_}
\lambda_{i} (T,\mu)=\exp\left(\frac{B_{i} \mu_{\mathtt{b}}+S_{i} \mu_{S}}{T} \right),
\end{equation}
where $B_{i} (\mu_{\mathtt{b}})$ and $S_{i} (\mu_{S})$ are baryon and strange quantum numbers (their corresponding chemical potentials) of the $i$-th hadron, respectively. From phenomenological point of view, the baryon chemical potential $\mu_{\mathtt{b}}$ can be related to $\sqrt{s_{\mathtt{NN}}}$ \cite{Tawfik:2013bza}
\bea
\mu_{\mathtt{b}} &=& \frac{a}{1+b \sqrt{s_{\mathtt{NN}}}},
\eea
where
$a=1.245\pm0.049~$GeV and $b=0.244\pm0.028~$GeV$^{-1}$. Then, the number density can be deduced as 
\begin{equation} \label{GrindEQ__6_}
n_i(T,\mu)=\sum_{i}\frac{\partial\, {\ln Z}_i(T,V,\mu)}{\partial \mu_{i}}=\sum_{i}\frac{g_{i}}{2\pi^{2}}\int_{0}^{\infty }\frac{p^{2} dp}{\exp\left[\frac{{\mu}_{i} - {\varepsilon}_{i}(p)}{T}\right] \pm 1}.
\end{equation}

The temperature $T$ and the chemical potential $\mu=B_{i} \mu_{\mathtt{b}}+S_{i} \mu_{S}+\cdots$ are related to each other and in turn each of them is related to $\sqrt{s_{\mathtt{NN}}}$ \cite{r4}. As an overall equilibrium is assumed, $\mu_{S}$ is taken as a dependent variable to be estimated due to the strangeness conservation, i.e. at given $T$ and $\mu_{\mathtt{b}}$, a value assigned to $\mu_{S}$ has to assure that $\langle n_S\rangle-\langle n_{\bar{S}}\rangle$ vanishes. Only then, $\mu_{S}$ joins $T$ and $\mu_{\mathtt{b}}$ in determining further thermodynamic calculations, such as the particle yields and the ratios. Chemical potentials for other quantum charges, such as electric change and isospin, can also be determined as functions of $T$, $\mu_{\mathtt{b}}$, and $\mu_{S}$ and each must fulfill the corresponding laws of conversation.

\subsection{Ultrarelativistic Quantum Molecular Dynamic (UrQMD) model}
\label{sec:urqmd}

The UrQMD hybrid model \cite{r26, r27, r28, r29, r30} assumes a non-homogeneous medium and combines various advantages of the hadronic transport theory and ideal fluid dynamics, as well. From the hydrodynamic evolution, the UrQMD hybrid model is conjectured to simulate almost the entire evolution of the heavy-ion collisions starting from a very early stage up to the final state of the particle production. Because of the different dynamics, symmetries, and effective degrees of freedom, for instance, the different stages are apparently distinguishable from each others. Therefore, different theoretical approaches must be utilized. On the other hand, the hydrodynamic models, which excellently describe the various stages of the nuclear collisions integrated in the UrQMD hybrid models furnish this with a unified framework, which in turn manifest the characteristics of the different stages. Furthermore, the large interaction rate and the microscopic Boltzmann transport models are also embedded. All these provide the UrQMD hybrid models with tools for a good description of the matter at low interaction rates.

For a hybrid event, we begin with UrQMD in cascade mode. Then, the nuclei are initialized and brought to collision. The hydrodynamic evolution starts when the nuclei with given radii $r$ have passed through each others. The corresponding temporal expansion $t_{hadro}=2r (\gamma^2-1)^{1/2}$, where $\gamma$ is the Lorentz factor. Assuming a local thermal equilibrium in each hydro cell, the so far produced particles are mapped onto a hydrodynamic grid. When the energy density of all hydro cells in a transverse slice with thickness $0.2~$fm drops to $20\%$ of the nuclear ground state density, the hydro degrees of freedom in this slice are finally mapped to particles (Cooper-Frye equation). 

The UrQMD hybrid model is a realistic and a well tested model for the background medium. It is a widely utilized event generator based on a large number of solutions of the Monte Carlo technique for a great number of partial differential equations giving the evolution of the phase-space densities and has a great number of unknown parameters which could be fixed from theoretical postulates and experimental results. Ingredients such as event-by-event fluctuations are included so that in this environment even the heavy quarks (charm and bottom) are placed in the nucleus-nucleus collision space-time-coordinates and allowed to propagate at each hydro time step in the hot medium by using the correspondent cell properties such as velocities, temperatures, and length of time-step (Langevin approach). At each time step and temperature, all particles are checked regarding being hadronized.

In the present work, we implement the UrQMD hybrid event generator version $3.4$ at various beam energies  $7$, $7.7$, $9$, $11$, $11.5$, $13$, $19$, $19.6$, $27$, $39$, $62.4$, $130$, and $200$ GeV. The fact that two different types of phase transition, namely, crossover and first order, are possible, allows us to run the UrQMD simulations for each of them, separately. With this we wanted to investigate which type of the phase transition matches well with the experimental results. To illustrate this point, we recall that the measurements, which are multiplicities of produced particles either in $4\pi$ detector acceptance and/or with limited rapidity, transverse momentum, etc. aren't providing any direct signature on the order of the phase transition. The latter was obtained in the first-principle lattice QCD and QCD-like approaches.

In case of crossover, equations of state of fluid dynamical evolution are utilized in the UrQMD hybrid model, while the MIT bag model and HRG approaches are used in the case of the first-order phase transition \cite{r30}. For the sake of completeness, we emphasize that there are two main differences between crossover and first-order phase transitions; the latent heat and the degrees of freedom. In the first-order phase transition both quantitie are larger than that in case of crossover. Furthermore, the crossover, as the same says, takes place, smoothly, i.e. a relative wide range of temperatures is needed to derive the QCD matter from hadron to parton phase or vice versa, while there is a prompt jump in case of first-order phase transition, i.e. the critical temperature becomes sharper \cite{r31}.
 
From an ensemble of at least $100,000$ events generated by the UrQMD hybrid model, we have calculated the ratios of different particle yields, at the energies $7$, $7.7$, $9$, $11$, $11.5$, $13$, $19$, $19.6~$, $27$, $39$, $62.4$, $130$, and $200$ GeV. In doing this, two types of phase transitions are taken into consideration; first order and crossover. The results are compared with the ones calculated from the HRG model and generated by CRMC EPOS $1.99$, are confronted to the available experimental results.

\subsection{Cosmic Ray Monte Carlo (CRMC) model}
\label{sec:crmc} 

EPOS is a parton model with many binary parton-parton interactions creating parton ladders. EPOS integrates energy-sharing for cross-section calculations, particle production, parton multiple scattering, outshell remnants, screening and shadowing via unitarization and splitting, and collective effects for dense media. Concretly, CRMC is a project incorporating an interface to the cosmic rays models for effective QCD-like models, Pierre Auger Observatory, and high-energy experiments such as NA61, LHCb, TOTEM, ATLAS, and CMS experiments. The cosmic ray models are build on top of the Gribov-Regge model such as EPOS $1.99$/LHC. CRMC offers a complementary description for the background including diffraction and also provides a common interface to access the output from various event generators for nuclear collisions. The interface is linked to a wide range of models, however, the unique focus is to models using simulations of extensive cosmic ray air showers such as qgsjet$01$ \cite{q0149, q0150}, qgsjetII \cite{qii51, qii52, qii53}, sibyll \cite{S54, S55, S56}, EPOS $1.99$ \cite{EE1, EE2}, QGSJET$01$, and SIBYLL$2.3$, at low energies. At high energies, EPOS $1.99$/LHC and QGSJETII v$03$ and v$04$ are the ones to be integrated in. 

EPOS is designed for the cosmic ray air showers and can be applied to pp- and AA-collisions at SPS, RHIC, and LHC energies. EPOS utilizes a simplified treatment of the interactions in the final state and can be used for minimum bias hadronic interactions in heavy-ion interactions \cite{EPOS}. It is worthy emphasizing that EPOS - even in the final state - doesn't cover the simulations for full hydro system. EPOS $1.99$/LHC is a model utilized in the present calculations, has a large number of parameters describing the essential quantities in physics and the phenomenological postulates. These can be fixed through experimental and theoretical assumptions. It is assumed that EPOS $1.99$/LHC draws a reasonable picture about the interactions between hadrons regarding the data generated from available experiments and event generators.  

In the present work, we use EPOS $1.99$ event generator at the energies $7$, $7.7$, $9$, $11$, $11.5$, $13$, $19$, $19.6$, $27$, $39$, $62.4$, $130$, and $200$ GeV. We have generated ensembles of at least $100,000$ events (at each of these energies). Proving the validity of EPOS $1.99$, we aim at predicting the corresponding particle ratios estimated at the freezeout parameters, which are novel predictions for the future facilities FAIR and NICA.

\section{Results and Discussion}
\label{results}  

The present analysis is focused on characterzing results measured at STAR BES-I energies \cite{Bzdak:2019pkr} and partially at the Alternating Gradient Synchrotron (AGS) \cite{r7} and the Superproton Synchrotron (SPS) \cite{r5} energies, as well, such as NA49 \cite{r33, r34, r35}, NA44 \cite{r33, r34, r35, r41, r51}, and NA57 \cite{r36}. A great part of this range of the beam energy shall be accessed by NICA and FAIR future facilities \cite{hanafy}. The calculations from CRMC EPOS $1.99$ (stars) \cite{EE1, EE2} and UrQMD hybrid model \cite{r26, r27, r28, r29, r30} with first-order (astrids) and crossover phase-transition (empty circles) are compared with the HRG calculations (solid curves) \cite{r2,r4,r5,r7,r8,r9,r10,r11,r12}, as well. 

\begin{figure}[htb]
\includegraphics[width=5cm]{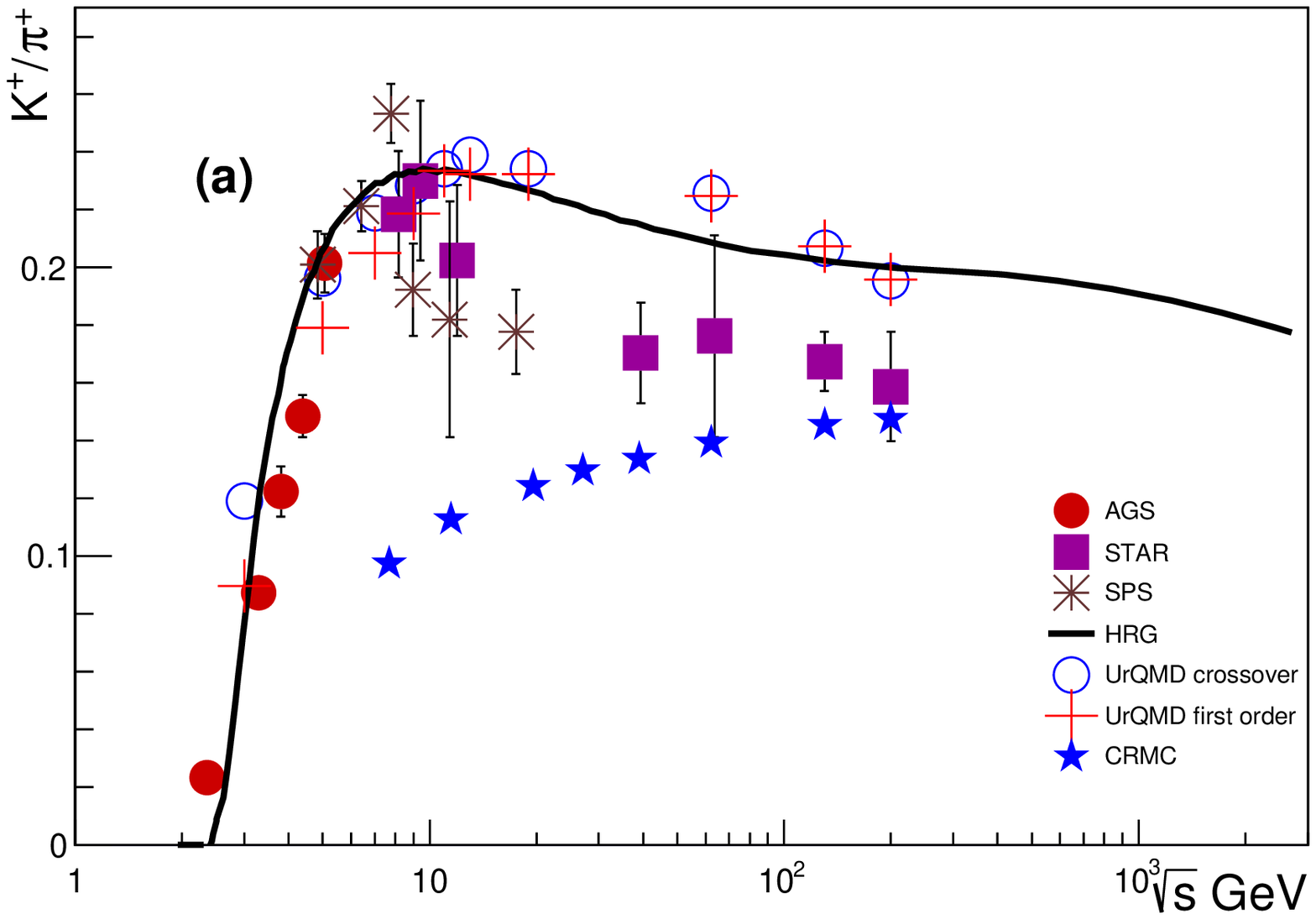}
\includegraphics[width=5cm]{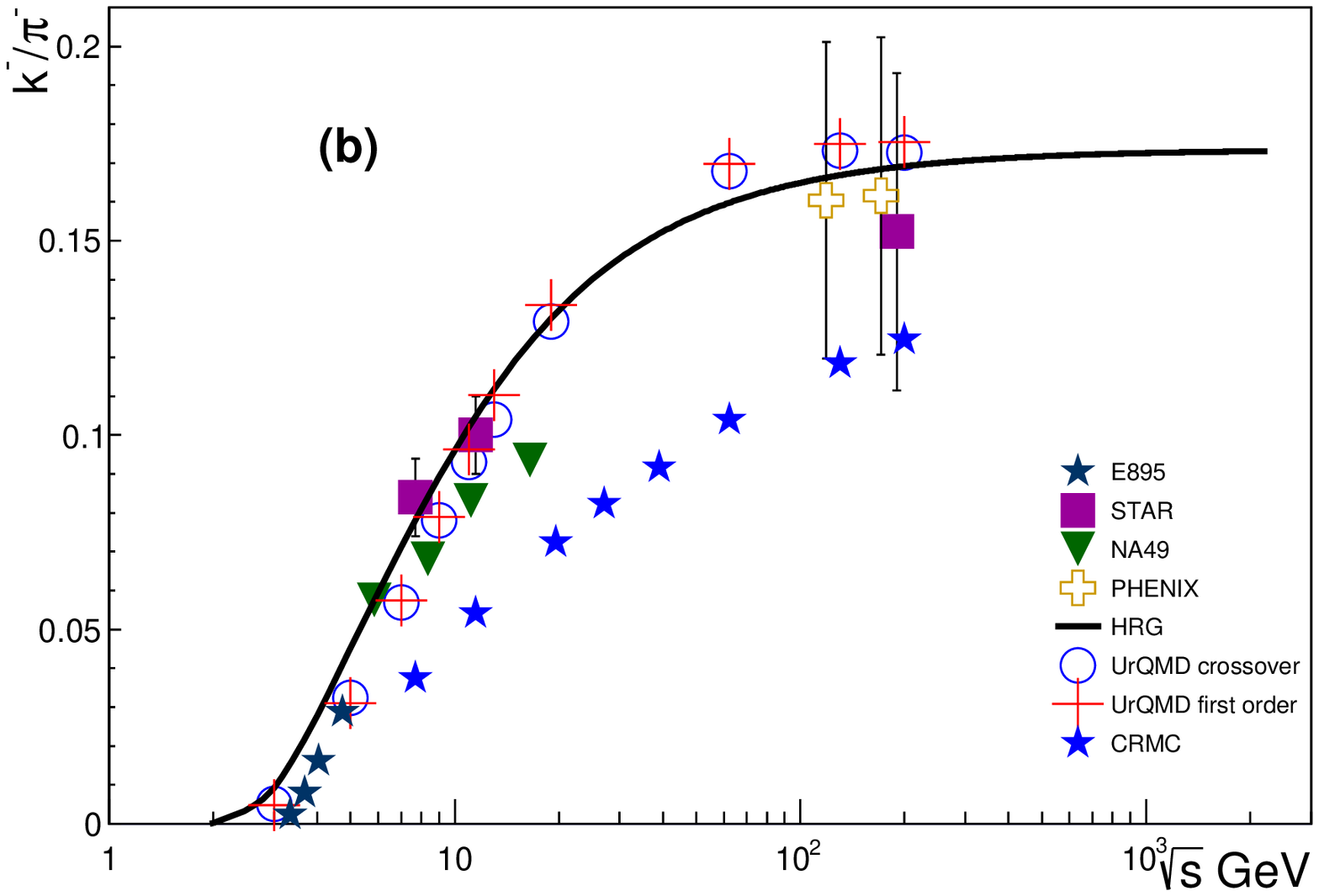} 
\includegraphics[width=5cm]{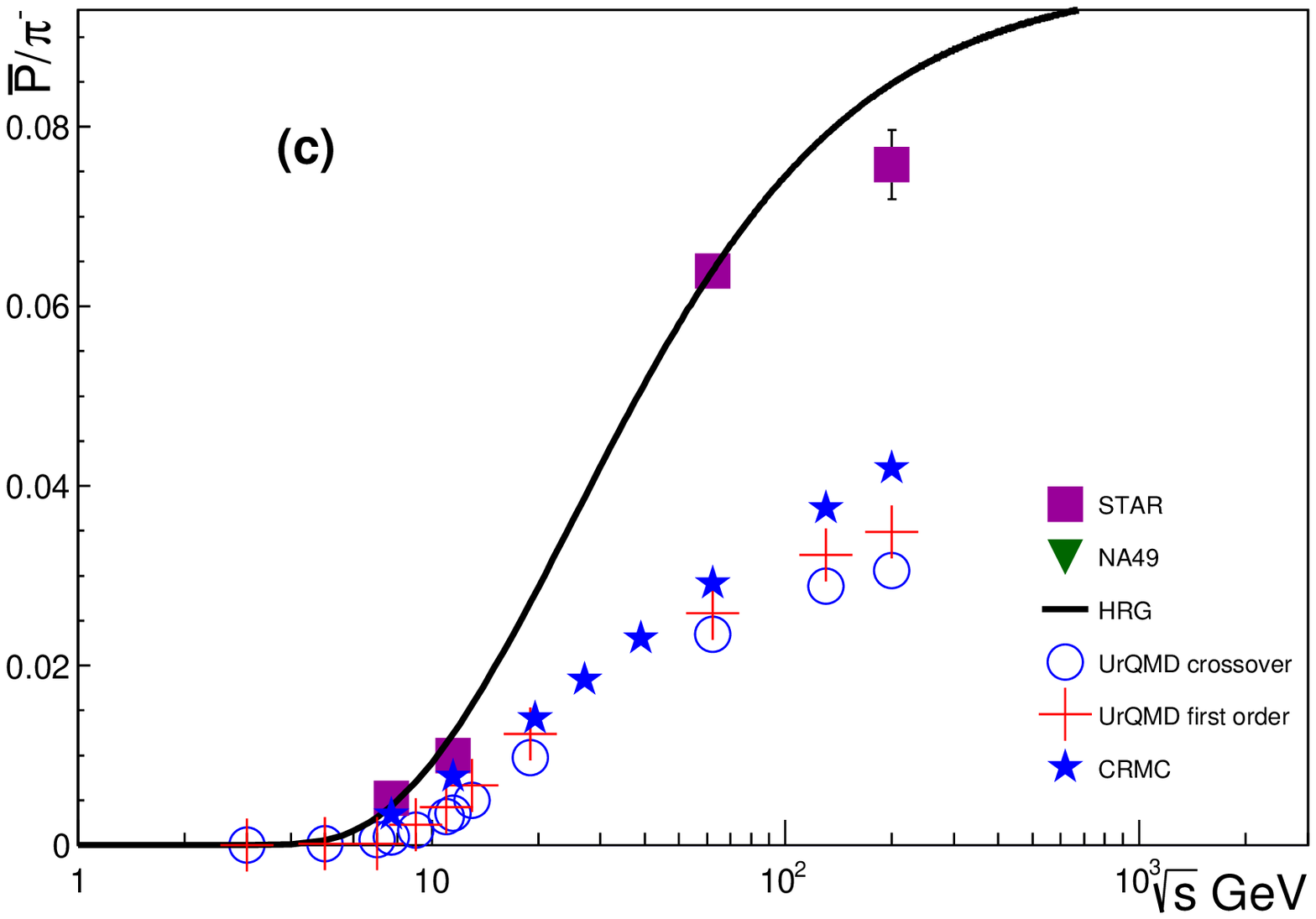} 
\includegraphics[width=5cm]{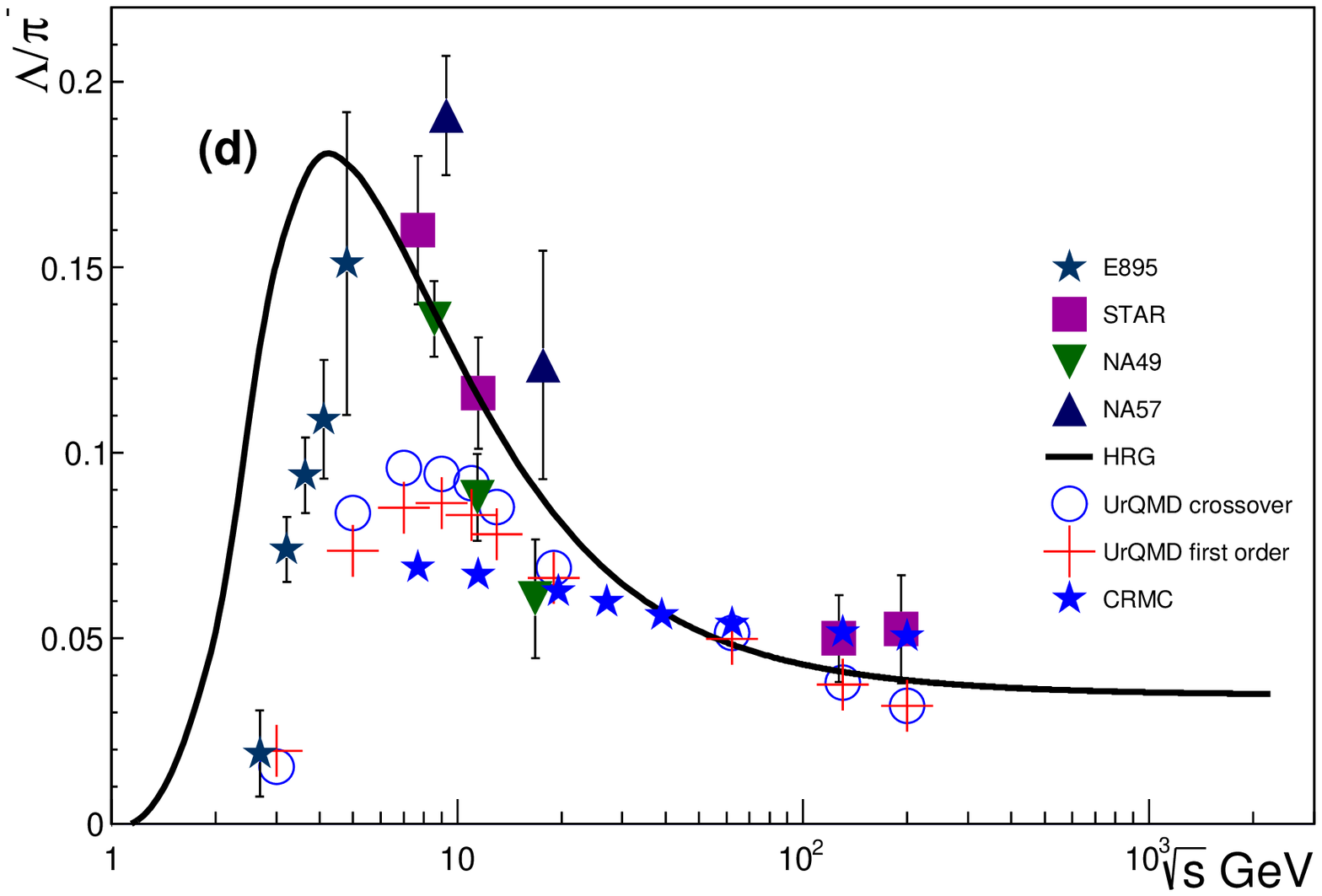}
\includegraphics[width=5cm]{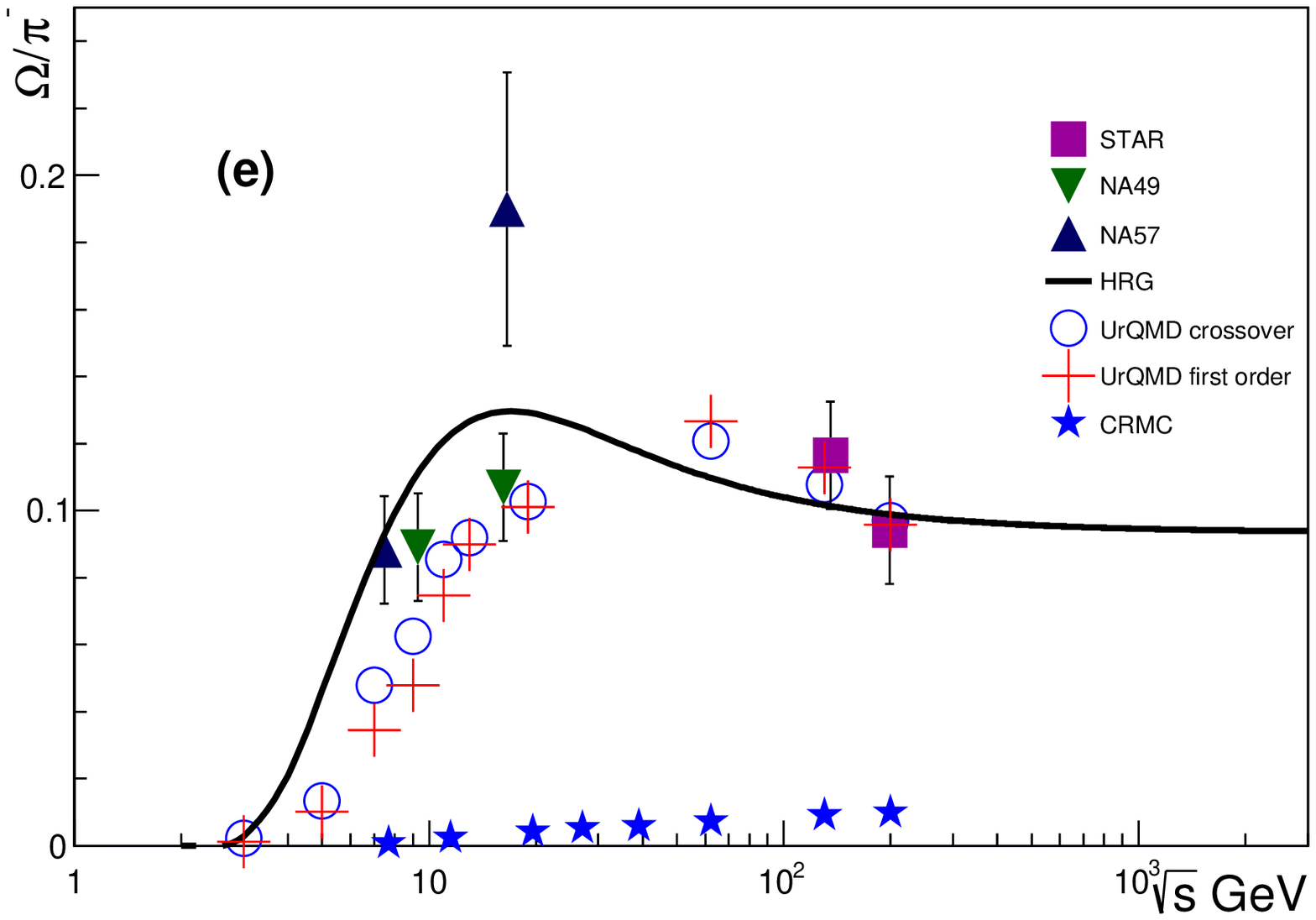} 
\includegraphics[width=5cm]{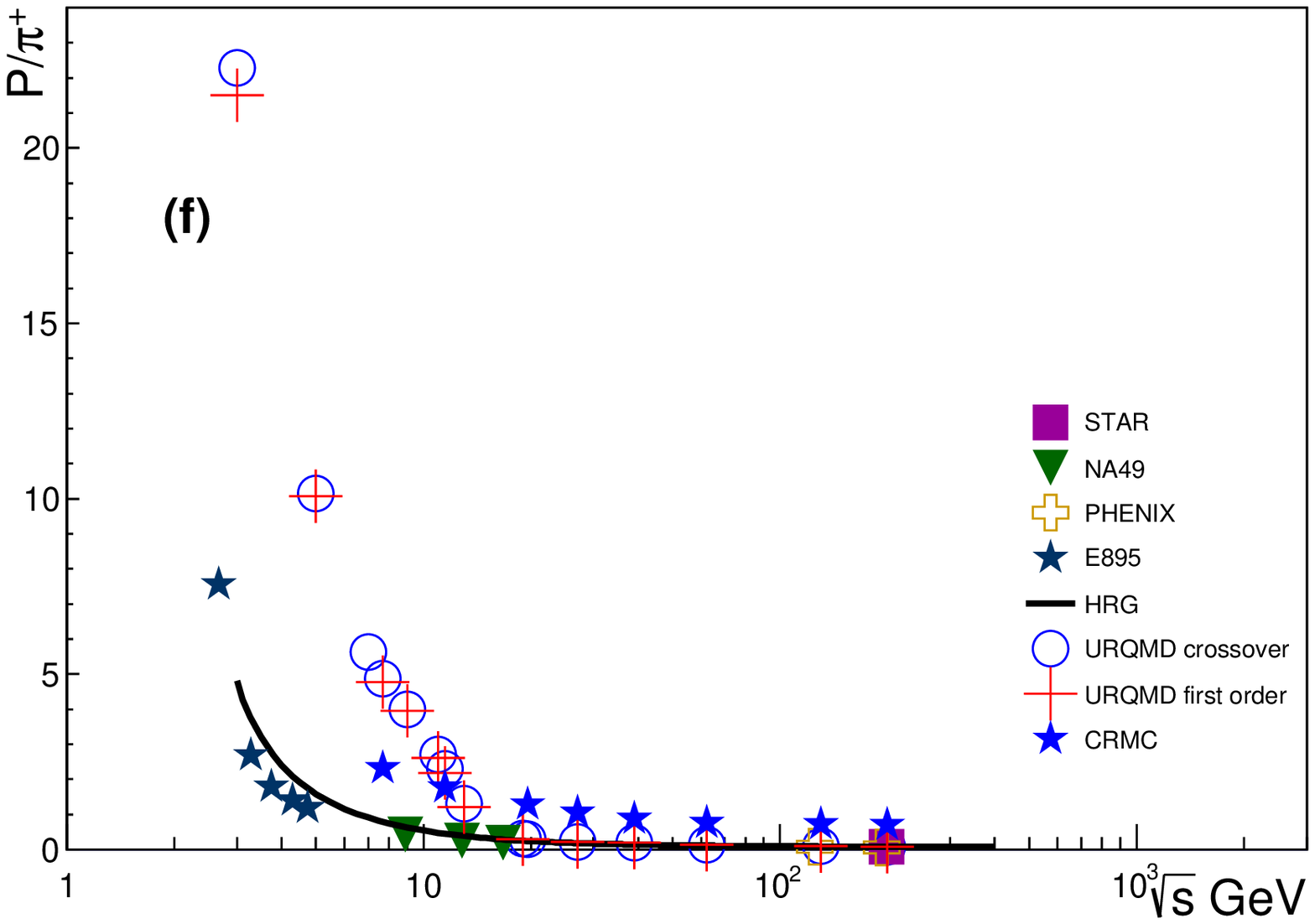} 
\caption{Dependence of various particle ratios calculated from CRMC EPOS $1.99$ (stars) \cite{EE1, EE2} and UrQMD hybrid model \cite{r26, r27, r28, r29, r30} with first-order (astrids) and crossover phase-transition (empty circles) is compared with HRG calculations (solid curves) \cite{r2,r4,r5,r7,r8,r9,r10,r11,r12} and different experimental results (symbols) on (a) $k^+/\pi^+$ \cite{r34, r35, r36}, (b) $\pi^-/K^-$ \cite{r34, r35, r36, r37}, (c) $\bar{p}/\pi^-$ \cite{r34, r35, r36, r38}, (d) $\Lambda/\pi^-$ \cite{r34, r35, r36, r37, r39, r40, r41}, (e)  $\Omega/\pi^-$ \cite{r34, r35, r36, r37, r39, r40, r41, r42} and (f) $p/\pi^+$ \cite{r34, r35, r36, r37, r39, r40, r41, r42}. }
\label{fig:one}
\end{figure}

\begin{figure}[htb]
\includegraphics[width=5cm]{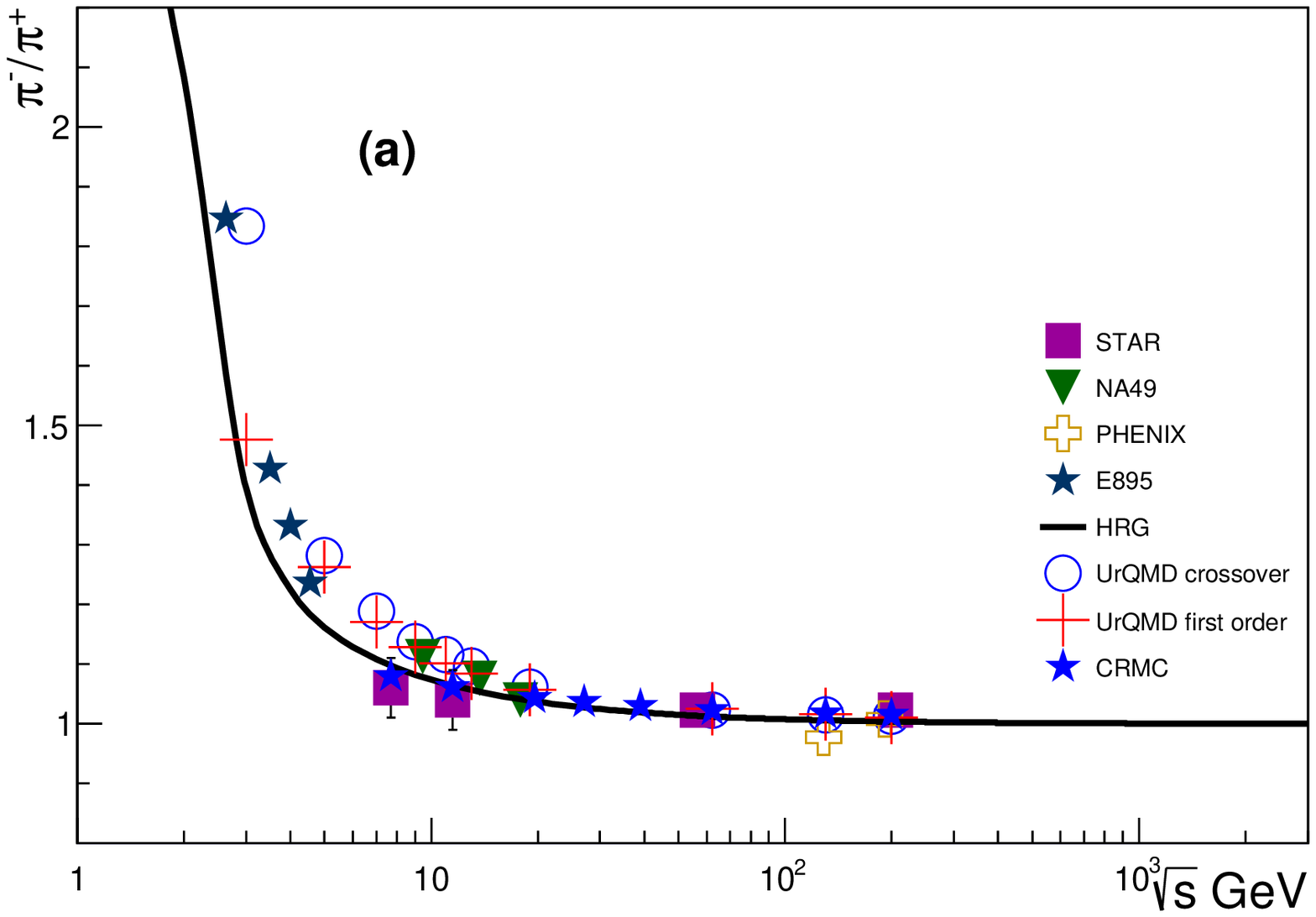}
\includegraphics[width=5cm]{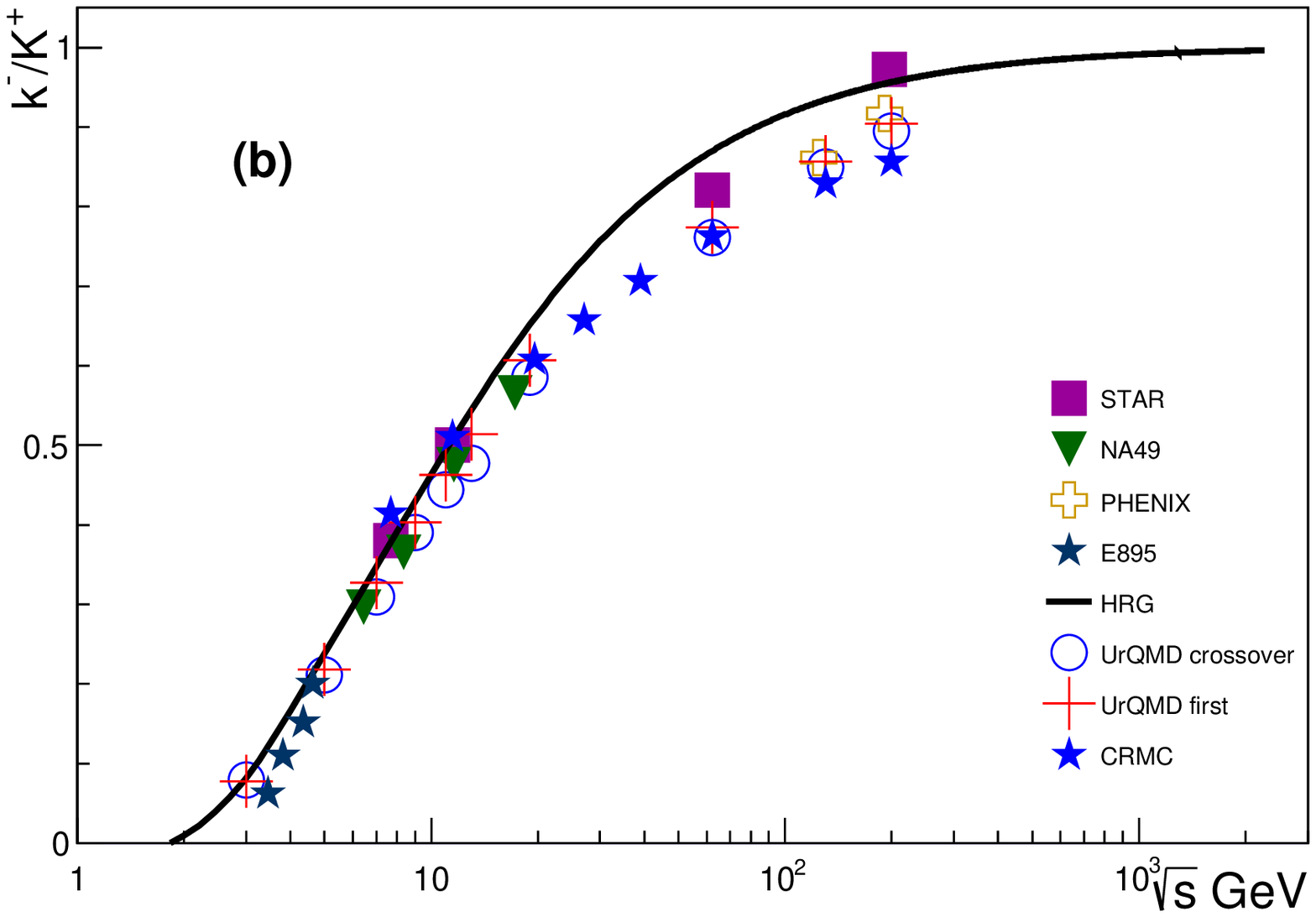}
\includegraphics[width=5cm]{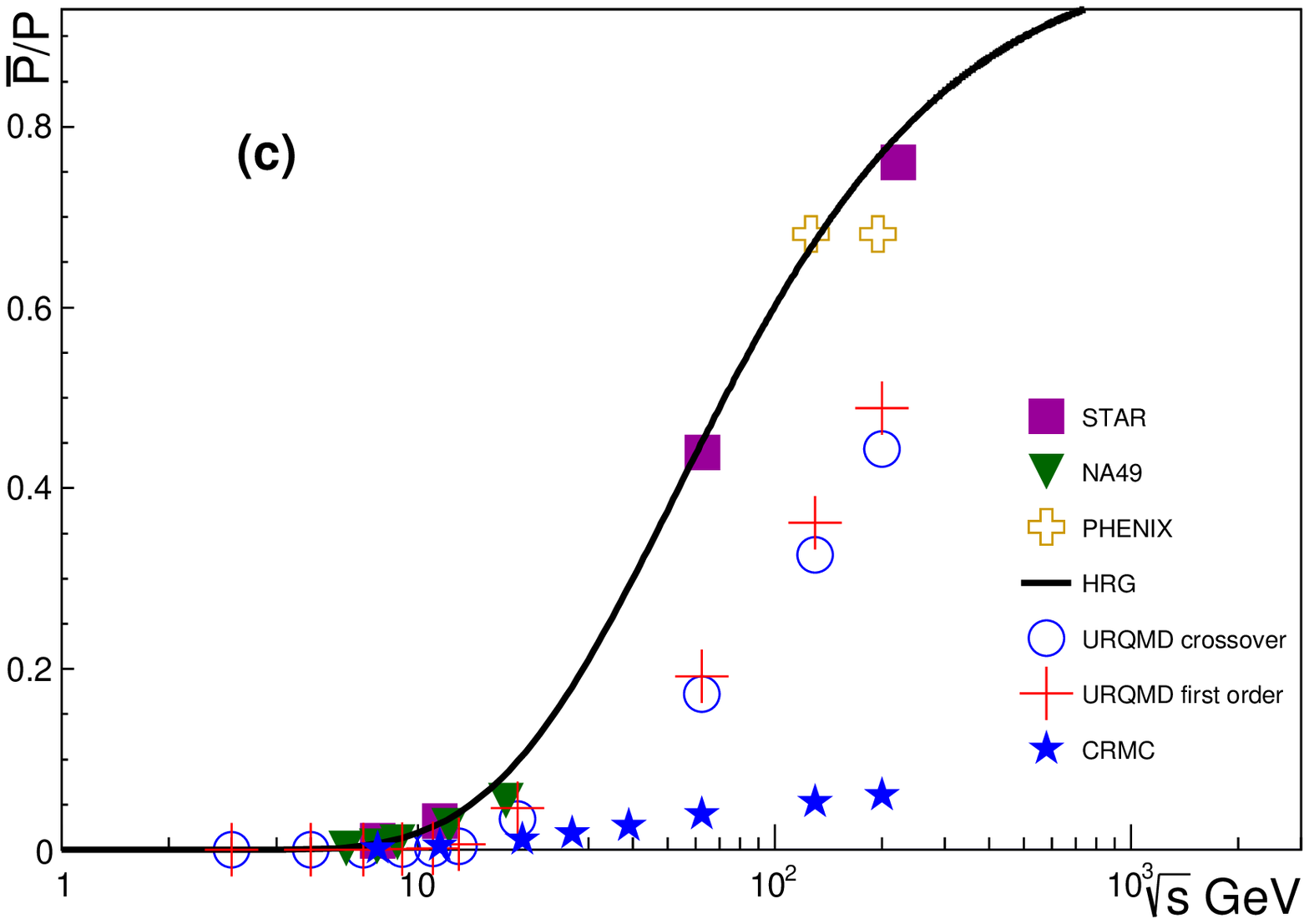} 
\includegraphics[width=5cm]{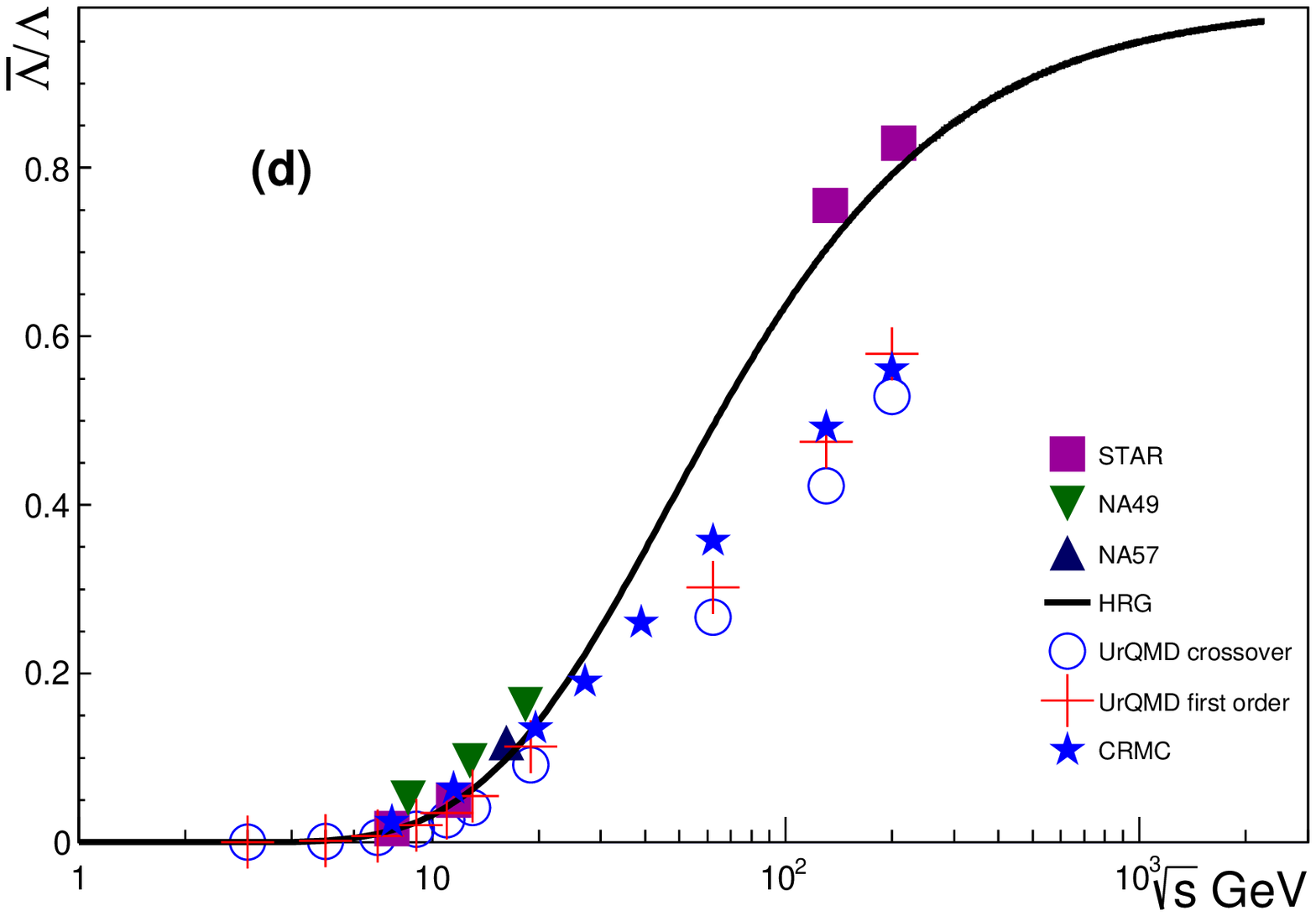} 
\includegraphics[width=5cm]{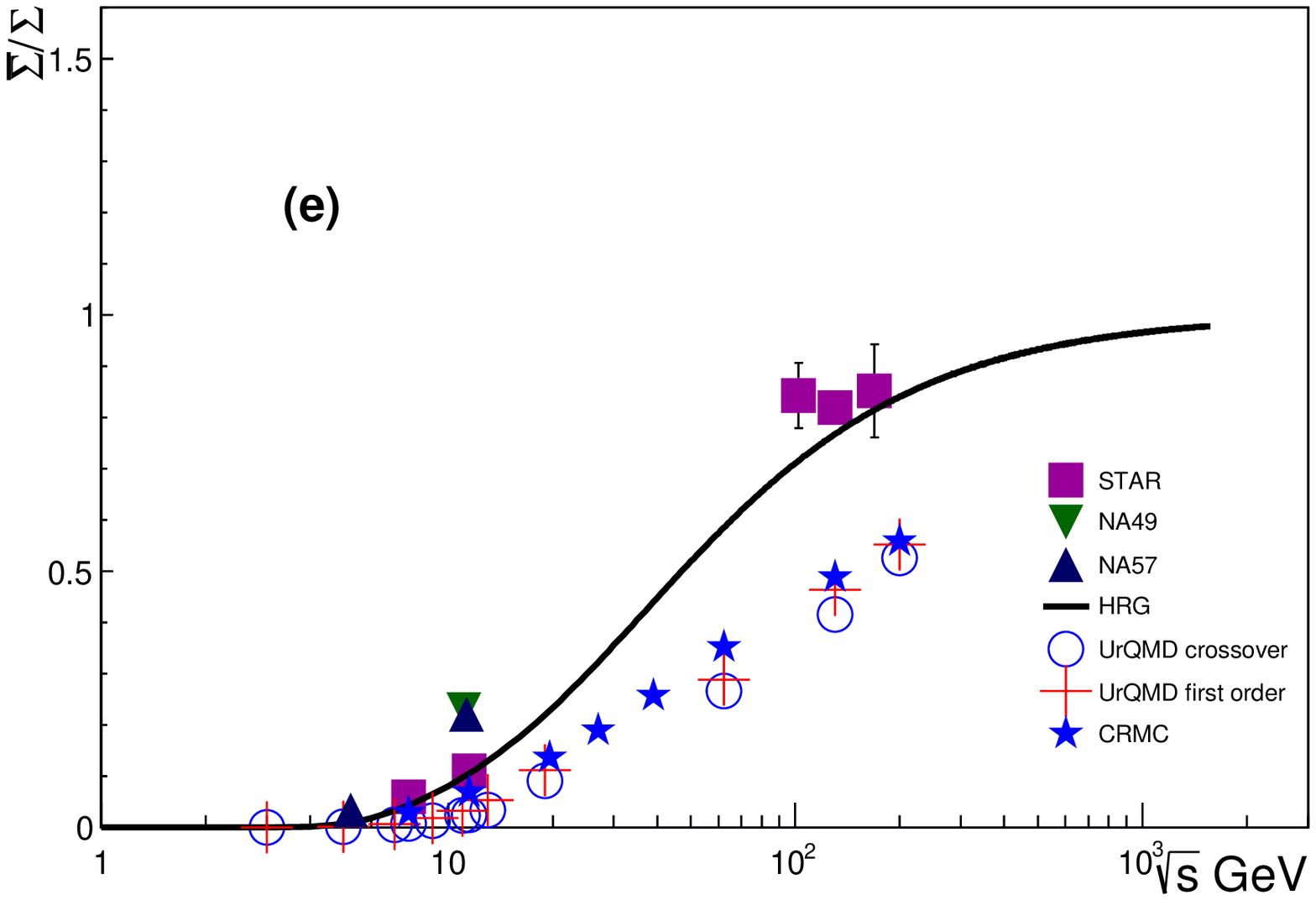}
\includegraphics[width=5cm]{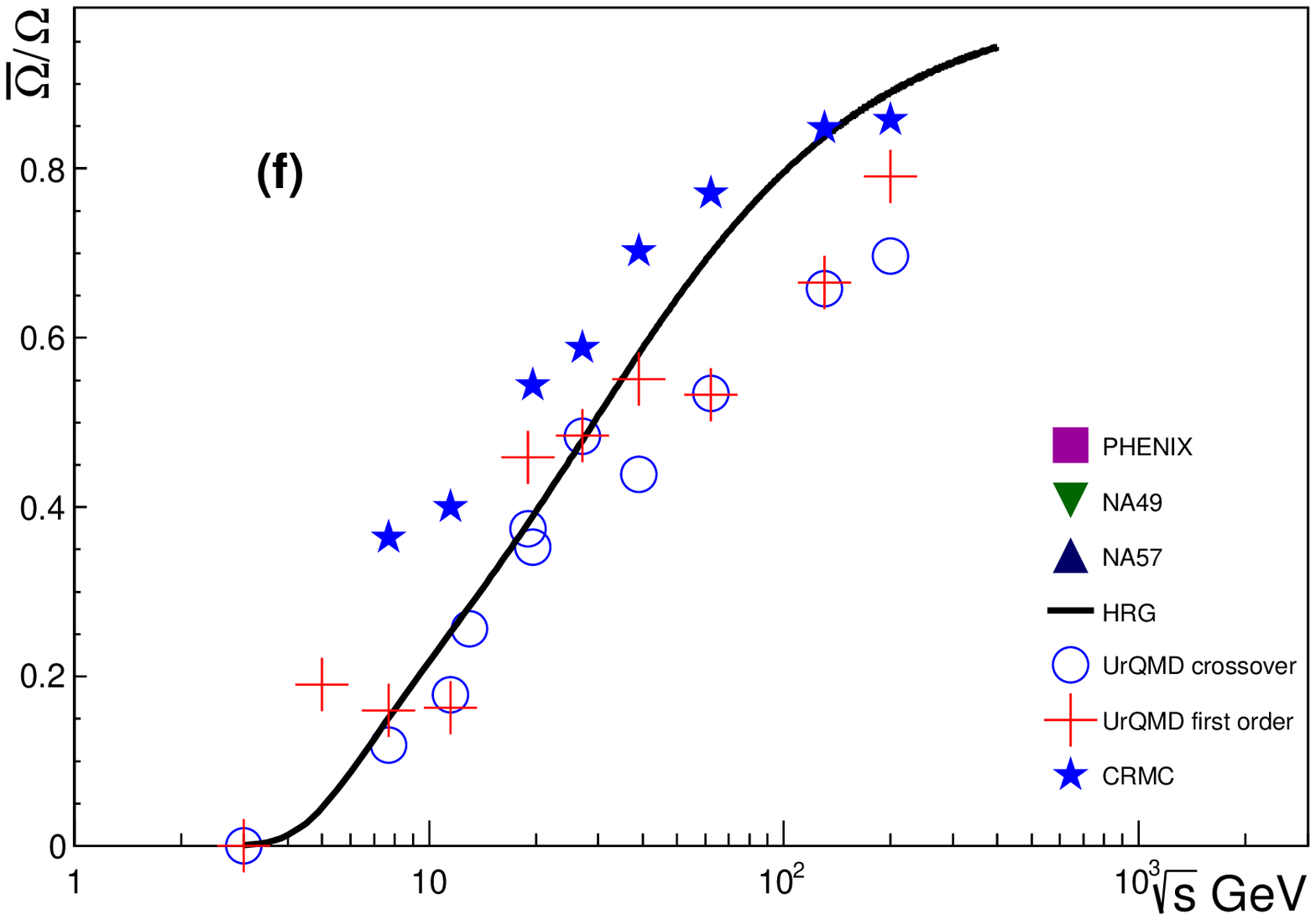}
\caption{The same as in Fig. \ref{fig:one} but here for particle-to-antiparticle ratios, (a) $\pi^-/\pi^+$ \cite{r34, r35, r36, r37}, (b) $K^-/K^+$ \cite{r34, r35, r36, r37}, (c) $\bar{p}/p$ \cite{r34, r35, r36, r38}, (d) $\bar{\Lambda}/\Lambda$ \cite{r34, r35, r36, r37, r39, r40, r41},  (e) $\bar{\Sigma}/\Sigma$ \cite{r34, r35, r36, r37, r39, r40, r41, r42} and (f) $ \bar{\Omega}/\Omega$ \cite{r34, r35, r36, r37, r39, r40, r41, r42}. } 
\label{fig:two}
\end{figure}

Figure \ref{fig:one} shows the energy dependence of the particle ratios  (a) $k^+/\pi^+$ \cite{r34, r35, r36}, (b) $\pi^-/K^-$ \cite{r34, r35, r36, r37}, (c) $\bar{p}/\pi^-$ \cite{r34, r35, r36, r38}, (d) $\Lambda/\pi^-$ \cite{r34, r35, r36, r37, r39, r40, r41, r42}, (e)  $\Omega/\pi^-$ \cite{r34, r35, r36, r37, r39, r40, r41, r42} and (f) $p/\pi^+$ \cite{r34, r35, r36, r37, r39, r40, r41, r42} measured at AGS, SPS and RHIC energies. Statistical ensembles of at least $100,0000$ events are generated by CRMC EPOS $1.99$ and UrQMD hybrid model at the energies $7$, $7.7$, $9$, $11$, $11.5$, $13$, $19$, $19.6~$, $27$, $39$, $62.4$, $130$, and $200$ GeV. In the UrQMD hybrid model, two types of the hadron-quark phase transition, namely crossover and first order, are taken into consideration. The results from the CRMC EPOS $1.99$ and the UrQMD hybrid model are compared with these measurements (symbols) and also with the HRG calculations (solid curves). 

Overall, there is a convincing agreement between the HRG model calculations on the given particle ratios and the experimental results and that from the event generators, CRMC EPOS $1.99$  and UrQMD. In some particle ratios, the UrQMD hybrid model seems to have a better agreement than that of the CRMC EPOS $1.99$. In case of the ratio $k^+/\pi^+$ \cite{r34, r35, r36}, we find that the HRG model agrees well with the UrQMD hybrid model and not with the CRMC EPOS $1.99$ event generator. The reproduction of the both experiment and event generator seems not possible. There is an overestimation by HRG observed at energies $\gtrsim 10~$GeV. CRMC EPOS $1.99$ give an opposite result, namely underestimation. For HRG and UrQMD hybrid model, a better situation is found in $\pi^-/K^-$ and $\Omega/\pi^-$, while CRMC EPOS $1.99$ again underestimates both particle ratios. We also find that the HRG model describes well the STAR measurements on $\bar{p}/\pi^-$, while the corresponding event-generator results are underestimating this particle ratio.  For $\Lambda/\pi^-$, the HRG reproduction of both experimental and simulation results is excellent, especially at energies $\gtrsim 10~$GeV. The HRG model describes well $p/\pi^+$. Here both event generators overestimate $p/\pi^+$, where CRMC EPOS $1.99$ is closer to HRG and experiments than UrQMD hybrid model .

Figure \ref{fig:two} shows the same as in Fig. \ref{fig:one} but here for the antiparticle-to-particle $\pi^-/\pi^+$ (a) \cite{r34, r35, r36, r37}, $K^-/K^+$ (b) \cite{r34, r35, r36, r37}, $\bar{p}/p$ (c) \cite{r34, r35, r36, r38},  $\bar{\Lambda}/\Lambda$ (d) \cite{r34, r35, r36, r37, r39, r40, r41}, $\bar{\Sigma}/\Sigma$ (e) \cite{r34, r35, r36, r37, r39, r40, r41, r42} and (f) $\bar{\Omega}/\Omega $ \cite{r34, r35, r36, r37, r39, r40, r41, r42}. We also compare the experimental results with the HRG calculations and the predictions from CRMC EPOS $1.99$ and UrQMD hybrid model.

Also here, we notice that while HRG agrees excellently with the experimental results.  While HRG reproduces well $\bar{p}/p$, $\bar{\Lambda}/\Lambda$ and $\bar{\Sigma}/\Sigma$, we find that CRMC EPOS $1.99$ and UrQMD hybrid model underestimate all these particle ratios. For the remaining particle ratios, we observe that the three data sets, namely the HRG model, the experiments and simulations agree well. Comparing to Fig. \ref{fig:one},  there is a better agreement for the antiparticle-to-particle ratios. This could be understood due to insubstantial fluctuations relative the mixed particle ratios.

\section{Conclusion}
\label{conc}

In the present work, the CRMC EPOS $1.99$ and the UrQMD hybrid model are utilized in generating statistical ensembles of $100,000$ events at $7$, $7.7$, $9$, $11$, $11.5$, $13$, $19$, $19.6~$, $27$, $39$, $62.4$, $130$, and $200$ GeV. At these energies $k^+/\pi^+$, $\pi^-/K^-$, $\bar{p}/\pi^-$, $\Lambda/\pi^-$, $\Omega/\pi^-$, $p/\pi^+$, $\pi^-/\pi^+$, $K^-/K^+$, $\bar{p}/p$, $\bar{\Lambda}/\Lambda$, $\bar{\Sigma}/\Sigma$, $ \bar{\Omega}/\Omega$ are determined. These results are then compared with various experiments at AGS, SPS and RHIC energies and with the HRG calculations. For the latter, the essential thermodynamic quantities, namely the temperature and the chemical potential, are determined at freezeout conditions, such as constant entropy density normalized to $T^3$.

For mixed particle ratios, $k^+/\pi^+$, $\pi^-/K^-$, $\bar{p}/\pi^-$, $\Lambda/\pi^-$, $\Omega/\pi^-$, $p/\pi^+$, $\pi^-/\pi^+$ there is a fair agreement between the HRG calculations and the experimental results and that from the event generators, CRMC EPOS $1.99$ and UrQMD hybrid model. We conclude that the UrQMD hybrid model seems to have a better agreement than that of the CRMC EPOS $1.99$. The so-called {\it horn} in $k^+/\pi^+$ isn't observed in both event generators, where CRMC EPOS $1.99$ largely underestimates this ratio as well as the $k^-/\pi^-$. While HRG reproduces well $\bar{p}/\pi^-$, both event generators largely underestimate this particle ratio, at all energies.

For $\Lambda/\pi^-$, the peak at energies $\sim 10~$GeV exists in all data sets. The HRG model reaches the same height as that of the experimental results, while both event generators produce smaller heights. In addition to these phenomenological observations, we find that the HRG model overestimates this ratio, at $\lesssim 7~$GeV.

We also found that in almost all particle ratios the two types of the phase transitions implemented in the UrQMD hybrid model are indistinguishable.  When focusing the comparison on CRMC EPOS $1.99$ and UrQMD hybrid model, we find that the results deduced from the UrQMD hybrid model agree excellently with the CRMC EPOS $1.99$. This isn't always the case, especially for $k^+/\pi^+$, $k^-/\pi^-$, $\Omega/\pi^-$,  $\bar{p}/\pi^+$, and $\bar{\Omega}/\Omega$. Apart from $\bar{\Omega}/\Omega$, CRMC EPOS $1.99$ seems to largely underestimate  $k^+/\pi^+$, $k^-/\pi^-$, $\Omega/\pi^-$,  and $\bar{p}/\pi^+$.
	
Last but not least, we firstly conclude that the HRG model reproduces excellently the experimental results on almost all particle ratios.  $k^+/\pi^+$ and $\Lambda/\pi^-$ are the partial exceptions. While HRG overestimates  $k^+/\pi^+$ , at $\sqrt{s_{\mathtt{NN}}}\gtrsim 10~$GeV , $\Lambda/\pi^-$ is overestimated, at $\sqrt{s_{\mathtt{NN}}}\lesssim 10~$GeV. Secondly, the reproduction of the HRG model for the event-generator results varies from particle ratio to another. Thirdly, the same conclusion can be drawn for both event generators and experimental results.




\end{document}